\documentclass[a4paper,11pt,fleqn]{article}
\usepackage{times}
\usepackage{multirow} 
\usepackage{hyperref}
\usepackage{breakurl}
\usepackage{datetime}

\begin{document}

\newcommand{\captionfonts}{\em}
\makeatletter  
\long\def\@makecaption#1#2{%
  \vskip\abovecaptionskip
  \sbox\@tempboxa{{\captionfonts #1: #2}}%
  \ifdim \wd\@tempboxa >\hsize
    {\captionfonts #1: #2\par}
  \else
    \hbox to\hsize{\hfil\box\@tempboxa\hfil}%
  \fi
  \vskip\belowcaptionskip}
\makeatother   

{\Large {\bf Evidence for periodicities in the extinction record?}}
\vspace*{0.65em}
\\
{\Large {\bf Response to Melott \& Bambach arXiv:1307.1884}}

\vspace*{0.5cm}
C.A.L.\ Bailer-Jones (calj@mpia.de), F.\ Feng (ffeng@mpia.de)\\
Max Planck Institute for Astronomy, Heidelberg, Germany\\
\today

\section*{Abstract}

In a recent paper, we applied time series analysis methods to study the possible influence of the solar motion through the Galaxy on terrestrial extinction  (Feng \& Bailer-Jones 2013). We drew conclusions about the relative probabilities of how well different models explain the geological extinction record. We found no strong evidence for either a periodicity in the extinction record or for a link between the solar orbit and this extinction record.  In response to this, Melott \& Bambach (2013) dedicate the majority of their article to criticizing our analysis. Their
main objection is to evidence-based model comparison. Here we deal with their criticisms one by one, and show that they are largely misplaced, leaving our conclusions unaffected. The bottom line is that although one can get a periodic model at some period to fit the extinction data, there are other models which explain the data better.

\section*{Introduction}

The article Melott \& Bambach (2013, hereafter MB) contains two unconnected parts. The first deals with a reassessment of their own earlier results. We will not discuss this part. The second and much longer part criticizes our recent paper (Feng \& Bailer-Jones, hereafter FBJ) in which we compare models of biological extinction variations in the fossil record. We deal with each of their criticisms below. These fall mostly into two categories. The first are assertions that our results may change if some assumptions are changed or the data or models are altered. Their second category of criticism centres on our use of the Bayesian evidence to do model comparison.  It appears MB have either not understood -- or do not accept -- this concept. We emphasize that the goal of FBJ was to do an independent comparison of different models of extinction variations (of which the periodic model MB focus on was just one). FBJ did not set out to establish whether some specific period is supported by the data.  These are quite different goals, a fact which MB may not have appreciated.

\section*{Response to individual sections}

We here respond to the individual criticisms, using bold to refer to the sections in MB. Some of these points were already discussed in FBJ, and the reader is referred to that article for more details and for references.

In {\bf section 3.2}, MB object to our use of the word ``biodiversity". This is a fair point: we should have used the term ``extinction rate'' instead in some places. Yet while we mention biodiversity in order to set the scene for our work, we clearly state that we deal only with the extinction rate in our analysis. It would be interesting to perform a similar analysis on actual ``biodiversity" data, but for this we need to use a  different time series modelling approach (see the end of section 5.2 in FBJ). 

In {\bf section 3.3}, MB make the point that the periodicities found in the extinction data may not exist in the full 550\,Myr of data we used, in particular not in the past 150\,Myr. It would indeed be interesting to repeat the analysis in FBJ for different blocks of the time scale. However, given that we are performing an independent model comparison, we would want to do this on various time blocks, and not just those supporting previous claims of certain periods.

Given that we started this work in 2012 using previously published data, we were not able to use a time scale published later in 2012 (MB's objection in {\bf section 3.4}). Whether adoption of this new timescale -- or some even better timescale in the future -- will make a significant difference to our analysis, MB cannot
claim without actually repeating our analysis.

In {\bf section 3.5}, MB quote Tukey as having said: ``Far better an approximate answer to the right question, which is often vague, than an exact answer to the wrong question, which can always be made precise." Leaving aside the issue of how one usefully answers vague questions, this quote makes the point we do in FBJ: Answer the right question! As we note in FBJ, and as many authors have argued in numerous publications, the frequentist p-value approach (which MB advocate) has serious deficiencies when it comes to model comparison. In particular, it is often used not to explicitly test the model of interest, but rather just to reject an alternative hypothesis. This is not always appreciated. The Bayesian evidence is preferred in this respect because it answers the question you actually want to address, namely ``is model X a better explanation of the data than model Y?" and not ``how often does some random model produce this particular period?".

It is worth noting that random data not infrequently show an apparently ``significant" period somewhere in the periodogram (see for example the simulations in section 4.2.1 of Bailer-Jones 2011). The Bayesian evidence is relatively insensitive to this. Contrary to what MB claim in {\bf section 3.5}, the evidence does take into account the strength of the periodogram peaks when it averages over the likelihoods. (This can be understood using the concept of the ``Occam factor", as discussed in texts on Bayesian methods.) And contrary also to what MB write, we do not just make assertions about a random model: it is one of the models we test.

It seems MB have not understood the goals of model comparison using the Bayesian evidence when they write (at the end {\bf section 3.5}): ``Furthermore they [FBJ] assert that they are testing the viability of periodic models by averaging over all such models. This of course dilutes any benefit of hitting the 'correct' model -- which is actually rather well-specified."  But this is exactly what you should do when you do not know in advance what the 'correct model' (here meaning 'parameter combination') should be. In other words, they object to using the evidence on the grounds that it is the average of the likelihood over the model parameters, and instead advocate using the highest likelihood. MB seem to be suggesting that we should not compare other models to a ``periodic model" but rather to a ``periodic model with period X and phase Y". We already discussed in FBJ why this is the wrong thing to do: a model with parameters fitted to the data invariably gives a higher likelihood than a model with unfitted parameters. If we are going to go down that path, then we should also fit our other models to the data and then compare like with like. But this corresponds to comparing the maximum likelihood of each model, which simply favours the more complex model. We discuss in detail -- and in several places (e.g. sections 1.2, 3.1, 5.2) -- in FBJ why this is wrong, so it is regrettable that MB still suggest one should do this.

Indeed, in their summary ({\bf section 5}), MB seem to reject the concept of using the Bayesian evidence entirely when they write ``they [FBJ] averaged over periodic models with all frequencies and phases, which via the Central Limit Theorem essentially randomizes the periodic models".  Although a little vague, this claim would have to be backed up: We are not aware of work showing that the Bayesian evidence is invalidated by the central limit theorem. The Bayesian evidence is a well established principle, present in text books and widely used in scientific work.  Marginalizing over parameters is a logical thing to do. The evidence of a periodic model is just the likelihood averaged over both period and phase. Similarly, the periodogram is the likelihood (as a function of period) averaged over the phase (Bretthorst 1998).

One point which we do make in FBJ is that the Bayesian evidence is sensitive to the adopted prior distribution over the parameters. This may be a problem when the prior cannot be specified exactly, which is often the case. One may expect that if we adopt very broad priors, then all models essentially end up with the same evidence. But this is not necessarily the case (not least because the prior must be normalized). Rather than arguing about this abstractly, we can examine how the evidence changes with the prior. We did this in FBJ (section 5.3) by adopting narrower priors, in particular one more concentrated around previously claimed values of the period. This does increase the evidence modestly, but not enough for the periodic model to be favoured over other models, thus refuting one of MB's main objections. For a fair comparison we should do the same with the other models by narrowing their priors around their best fitting values. In the limit of infinitesimally narrow priors, we are then just comparing the maximum likelihoods, which, as we have already noted, is invalid. One reason to use the Bayesian evidence is that it takes into account the different model complexities for you.

We are not claiming that there are no issues with Bayesian model comparison, just that MB's criticisms thereof are either untenable or were addressed in FBJ. One issue which MB do not raise, but which we did highlight in FBJ and tested empirically there (e.g.\ Fig. 12), is that the data may not discriminate very well between the models. That is why we do not rule out periodic models in FBJ. Our conclusion was rather that they do not need to be invoked to explain the data. We emphasize again -- as we did in FBJ -- that any analysis may only draw conclusions about models which it explicitly tests.

MB then move on to other topics by suggesting, in {\bf section 3.6}, that the time of some events may actually be systematically shifted to the end of each stage. As the evidence is a calculation over all phases, our analysis accommodates a common shift. We did not test the effect of applying different shifts to each event independently. 
But without a specific test neither we (nor MB) can know whether this will change our conclusions.

In {\bf section 3.7}, MB claim that the uncertainties of the extinction dates may be less than what we adopted. They first suggest that these uncertainties should be less than about 1\,Myr. In fact, the uncertainties we adopted for the B18 data are of this order: as stated at the end of section 2.1 in FBJ, we use $\sigma=d/\sqrt{12}$ as the uncertainty (Gaussian standard deviation), where $d$ is the substage duration in Table 1 in FBJ. $\sigma$ therefore ranges from 0.6\,Myr to 2.4\,Myr for all events, with only 3 of 18 events having $\sigma$ larger than 1.4\,Myr. 
Moreover, as reported in section 5.3 of FBJ, we tested the impact of reducing these uncertainties, and found that it hardly changes the evidence. 

In {\bf sections 3.8 and 3.9}, MB point out that we have ignored periodicities discovered in other work. Yet this is precisely the point of doing an independent model comparison.  It seems such such independent tests may not be welcome when we read, in {\bf section 3.8}: ``Two additional statistical tests are possible, as shown [in] those papers, in which the question is formulated as finding agreement of a period with that already found in independent data. FB-J chose not to do such tests."

MB reference their earlier works where they find their period in other data sets. We agree that it would be interesting and useful to extend our analysis to these data sets, but there is only so much one can or should do in a single paper. In FBJ we introduced a new approach into this field, which we would be happy to extend in subsequent work.

MB's periodogram shown in {\bf section 3.10} does not add to their arguments.  Indeed, it suggests that small systematic errors in the dating would lead one to disregard the period.  But it also distracts from the main point: parameter estimation is not model comparison. We are interested in whether the periodic model is better than other models, not whether we can find one specific period which fits the data.

\section*{Summary}

In summary, MB's criticisms fall into two categories:

First, they suggest that introducing various changes in the data or models may change our conclusions.  We performed some such sensitivity tests in FBJ (and more not published) and found they made little difference. We could of course do more, and some of MB's suggestions are useful. Yet in the absence of actual investigations, MB's claims that further perturbations could change our conclusions are just assertions.

MB's second set of criticisms arises from either not understanding or not accepting the concept of (evidence-based) model comparison. In particular, MB object that model comparison does not take into account previous evidence for a period.  But if one assumes from the beginning that the data should show some sinusoidal period (27\,Myr, 62\,Myr, etc.), there is little point comparing various periodic and non-periodic models, which was the goal of our work. We chose to perform an independent study which asks ``which of a set of models best describes the data?", without regard to previous claims of periodicity and without comparing models tuned to such periods. 
One may not agree with this goal, but it is not a valid criticism of our work.

There are of course aspects of FBJ which could be improved or extended upon.  We outlined some in the original paper, and MB have suggested some additional tests.  But beyond this, MB's criticisms are misplaced and do not lead us to doubt the conclusions of our work.

\section*{References}

Bailer-Jones C.A.L., 2011, \href{http://adsabs.harvard.edu/abs/2011MNRAS.416.1163B}{MNRAS 416, 1163}\\
Bretthorst G.L., 1998, \href{http://bayes.wustl.edu/glb/book.pdf}{{\em Bayesian spectrum analysis and parameter estimation}},\\
\hspace*{1em}Lecture notes in statistics vol.\ 48, Springer\\
Feng F., Bailer-Jones C.A.L., 2013, \href{http://adsabs.harvard.edu/abs/2013ApJ...768..152F}{ApJ 768, 152}\\
Melott A.L., Bambach R.K., 2013, \href{http://arxiv.org/abs/1307.1884v1}{arXiv:1307.1884v1}

\end{document}